\magnification=\magstep1
\def\meno{\medskip\noindent}

\def\disleft#1:#2:#3\par{\par\hangindent#1\noindent
			 \hbox to #1{#2 \hfill \hskip .1em}\ignorespaces#3\par}

\def\boxit#1{\vbox{\hrule\hbox{\vrule\kern3pt \vbox{\kern3pt#1\kern3pt}\kern3pt\vrule}\hrule}}

\font\Bigmath=cmsy10 scaled \magstep2

\def\diamondplus{\mathrel{%
  \ooalign{\raise.29ex\hbox{$\scriptscriptstyle+$}\cr\hss$\diamond$\hss}}}
  
\def\diamondplustwo{\mathrel{%
  \ooalign{$+$\cr\hss\lower.255ex\hbox{\Bigmath\char5}\hss}}}

\font\eightrm=cmr8

-lmtk10

\def\display#1:#2:#3\par{\par\hangindent #1 \noindent
	\hbox to #1{\hfill #2 \hskip .1em}\ignorespaces#3\par}

\def\xskip{\hskip 7pt plus 3pt minus 4pt}
\newdimen\algindent
\newif\ifitempar \itempartrue 
\def\algindentset#1{\setbox0\hbox{{\bf #1.\kern.25em}}\algindent=\wd0\relax}
\def\algbegin #1 #2{\algindentset{#21}\alg #1 #2} 
\def\aalgbegin #1 #2{\algindentset{#211}\alg #1 #2} 
\def\alg#1(#2). {\medbreak 
  \noindent{\bf#1}({\it#2\/}).\xskip\ignorespaces}
\def\algstep#1.{\ifitempar\smallskip\noindent\else\itempartrue
  \hskip-\parindent\fi
  \hbox to\algindent{\bf\hfil #1.\kern.25em}%
  \hangindent=\algindent\hangafter=1\ignorespaces}
\def\thbegin#1. {\noindent {\bf#1.}\xskip\ignorespaces}
\def\proof.{\medbreak\noindent{\it Proof.}\xskip\ignorespaces}
\def\solution.{\medbreak\noindent{\it Solution.}\xskip\ignorespaces}

\input amssym.def
\input amssym.tex

\newdimen\rotdimen
\def\vspec#1{\special{ps:#1}}
\def\rotstart#1{\vspec{gsave currentpoint currentpoint translate
   #1 neg exch neg exch translate}}
\def\rotfinish{\vspec{currentpoint grestore moveto}}
%
%
\def\rotr#1{\rotdimen=\ht#1\advance\rotdimen by\dp#1%
   \hbox to\rotdimen{\hskip\ht#1\vbox to\wd#1{\rotstart{90 rotate}%
   \box#1\vss}\hss}\rotfinish}
%
%
\def\rotl#1{\rotdimen=\ht#1\advance\rotdimen by\dp#1%
   \hbox to\rotdimen{\vbox to\wd#1{\vskip\wd#1\rotstart{270 rotate}%
   \box#1\vss}\hss}\rotfinish}%
%
%
\def\rotu#1{\rotdimen=\ht#1\advance\rotdimen by\dp#1%
   \hbox to\wd#1{\hskip\wd#1\vbox to\rotdimen{\vskip\rotdimen
   \rotstart{-1 dup scale}\box#1\vss}\hss}\rotfinish}%
%
%
\def\rotf#1{\hbox to\wd#1{\hskip\wd#1\rotstart{-1 1 scale}%
   \box#1\hss}\rotfinish}%

\def\Cook{1}
\def\Levin{2}
\def\Karp{3}
\def\Hema{4}
\def\Gasa{5}
\def\Davis{6}
\def\Loge{7}
\def\Marques{8}
\def\Selman{9}
\def\Garey{10}

\parskip 2pt

\centerline{\bf Understanding SAT is in P}
\centerline{Alejandro S\'anchez Guinea\footnote{*}{\eightrm University of Luxembourg, 
\vskip .01cm Email: ale.sanchez.guinea@gmail.com or alejandro.sanchezguinea@uni.lu}
}

\bigskip
\bigskip
{\narrower\smallskip\noindent
{\bf Abstract.}\enspace
We introduce the idea of an understanding with respect to a set of clauses as
a satisfying truth assignment explained by the contexts of the literals in the clauses. 
Following this idea, we present
a mechanical process that obtains, if it exists, an understanding
with respect to a 3$\,$SAT problem instance based on the contexts of each literal in 
the instance, otherwise it determines that none exists. 
We demonstrate that our process is correct
and efficient in solving~3$\,$SAT.
\smallskip}

{\baselineskip14pt

\meno
Satisfiability (SAT, for short) is regarded as one of the most fundamental computational problems.
In the 1970's, when the class NP of problems was first defined~[\Cook,~\Levin,~\Karp], both SAT and 
its special case 3$\,$SAT were among the first problems shown to be NP-complete in~[\Cook].
This highlighted the importance of 3$\,$SAT and SAT since, by the definition of 
NP-completeness, 
if a polynomial-time algorithm exists for 3$\,$SAT or SAT then all NP-complete problems, 
and indeed all problems in NP, can be solved efficiently.
To date, however, no polynomial-time algorithm has been found for any
of the NP-complete problems, which among other reasons has led to the widely
accepted belief that no such algorithm exists~[\Hema,~\Gasa].

Although no polynomial-time algorithm has been found for SAT or 3$\,$SAT,
remarkable improvements in terms of efficiency have been achieved 
throughout the years. This has been accentuated recently (starting in the 1990's)
in the form of so called SAT solvers, 
which are practical procedures 
for SAT able to handle large instances considerably fast.
Modern SAT solvers can be divided into two groups:
{\it a)} complete solvers, mainly based on the backtracking
search procedure of Davis-Putnam-Logemann-Loveland (DPLL) algorithm~[\Davis,~\Loge] and
Conflict-Driven Clause Learning (CDCL) algorithm~[\Marques],
which are meant to always provide the correct solution given enough time;
and {\it b)} incomplete solvers, mostly based on stochastic local search~[\Selman],
which at the expense of statistically minimal errors seek to produce a quick answer.

As mentioned above, in spite of great advances, until now no algorithm
has been proposed to solve SAT or 3$\,$SAT in polynomial time.
In this paper we propose an algorithm that achieves this for 3$\,$SAT.
We introduce the idea of an understanding with respect to a set of clauses as
a satisfying truth assignment explained by the contexts of the literals in the clauses,
where the key point is the use of contexts which allow to construct the assignment
without searching (locally or systematically) the space of potential solutions.
Following this idea, our algorithm obtains, if it exists, an understanding
with respect to a~3$\,$SAT problem instance based on the contexts of each literal in 
the instance, otherwise it determines that none exists.

The outline of the paper is as follows. In $\S$~0 we recall definitions
on 3$\,$SAT. 
In $\S~1$ we present the idea introduced in this paper, including 
definitions, lemmas, and algorithms that lead to our main algorithm.
Finally, in $\S~2$ we present the analysis of our main algorithm in terms of correctness
and asymptotic time complexity.

\bigskip\noindent
{\bf 0. Preliminaries.}\quad
We recall some definitions from~[\Garey]. 

Let \hbox{$X=\{x_{1}, x_{2}, \dots, x_{m}\}$} be a set of Boolean {\sl variables}.
A {\sl truth assignment} for $X$ is a function $\alpha\colon X\to\{0, 1\}$.
If $\alpha(x)=1$ we say that $x$ is ``true'' under $\alpha$; 
if $\alpha(x)=0$ we say that $x$ is ``false''.
If $x$ is a variable in $X$, then $x$ and $\overline{x}$ are {\sl literals}
over $X$. We say that $\overline{x}$ is the {\sl negation} of $x$ and
$x$ is the negation of $\overline{x}$. 
The literal $x$ is true under $\alpha$ \hbox{if and only if} the variable $x$ is
true under $\alpha$; the literal $\overline{x}$ is true \hbox{if and only if}
the variable $x$ is false.
 
A {\sl clause} over $X$ is a set of literals over $X$, such as 
$\{x_{1}, \overline{x}_{3}, x_{8} \}$. It represents the disjunctions of 
those literals and is {\sl satisfied} by a truth assignment \hbox{if and only if}
at least one of its members is true under that assignment.
The clause above will be satisfied by $\alpha$ unless
$\alpha(x_{1})=0$, $\alpha(x_3)=1$, and $\alpha(x_{8})=0$. 
A collection $\Phi$ of clauses over $X$ is {\sl satisfiable} \hbox{if and only if}
there exists some truth assignment for $X$ that simultaneously satisfies all
clauses in $\Phi$. Such a truth assignment is called
a {\sl satisfying truth assignment} for $\Phi$.

The {\sl 3-satisfiability (3$\,$SAT)} problem is specified as follows:

Given a collection $\Phi=\{\varphi_{1}, \varphi_{2}, \dots, \varphi_{m}\}$
of clauses on a finite set $X$ of variables such that~$|\varphi_{i}|=3$
for~$1\leq i \leq m$. Is there a truth assignment for $X$ that satisfies
all clauses in~$\Phi$$\,$?

\bigskip\noindent
{\bf 1. Idea.}\quad
Let $\Phi$ be an instance of the 3$\,$SAT problem on a finite set $X$
of variables, and let $L$ be the set of all literals over $X$.
An {\sl understanding} for $L$ is a function
$\tilde{u}\colon L\to\{t, f, \varepsilon\}$.
For any literal $\lambda$ in $L$, if $\tilde{u}(\lambda)=t$ we say that 
$\lambda$ is ``true'' under $\tilde{u}$;
if $\tilde{u}(\lambda)=f$ we say that $\lambda$ is ``false''; 
and if $\tilde{u}(\lambda)=\varepsilon$ we say that $\lambda$ is ``free''.

The literal~$x$ is true and its negation $\overline{x}$ is false 
under $\tilde{u}$
\hbox{if and only if} the variable $x$ is true under $\alpha$;
the literal $\overline{x}$ is true
and its negation $x$ is false under $\tilde{u}$
\hbox{if and only if} the variable $x$ is false under $\alpha$;
the literal $x$ is free  
and its negation $\overline{x}$ is free under $\tilde{u}$
\hbox{if and only if} the variable $x$ is unassigned.

Let $\varphi\colon\{l_{1}, l_{2}, l_{3}\}$ be a clause in $\Phi$, 
where $l_{1}$, $l_{2}$, and $l_{3}$ are all distinct literals. We assume 
that $\varphi$ is satisfied.
If we focus on one of the literals in $\varphi$, say $l_{1}$, we say that
the {\sl context} of $l_{1}$ in $\varphi$ is the set of literals that
appear in $\varphi$ that are different from $l_{1}$, i.e.,~$\{l_{2}, l_{3}\}$. 
We call {\sl concept} to a context in which
its literals are interpreted according to a particular understanding.
Thus, the concept of $l_{1}$ in $\varphi$, interpreted according to some 
understanding~$\tilde{u}$, is denoted as
${\cal C}\colon\{\tilde{u}(l_{2}), \tilde{u}(l_{3})\}$.

Based on the elements in the codomain of a function of understanding, 
the possible combinations of elements, under some understanding $\tilde{u}$, 
in some concept~${\cal C}$ are: 
\smallskip
\settabs\+\indent&Horizontal lists\quad\quad\quad&\cr 
\+&\quad \item{\it i)} both literals free; 
& \quad\qquad\qquad \item{\it iv)} one literal free and the other true;\cr
\+&\quad\item{\it ii)} both literals true; 
& \quad\qquad\qquad \item{\it v)} one literal true and the other false;\cr
\+&\quad\item{\it iii)} both literals false; 
& \quad\qquad\qquad \item{\it vi)} one literal free and the other false.\cr
\smallskip

\noindent
We say that a concept as in~{\it (i)}, {\it (iii)}, or~{\it (vi)} is of type~${\cal C}^{+}$
and a concept as in~{\it (ii)}, {\it (iv)}, or~{\it (v)} is of type~${\cal C}^{*}$.

Let~$\phi$ be a set of clauses, subset of $\Phi$, that are assumed to be satisfied.
Let~$\widetilde{{\cal C}}$ be the set of all concepts of literals in clauses of~$\phi$
interpreted according to an understanding~$\tilde{u}$.
Let $\lambda$ be any literal that appears in one or more clauses in~$\phi$ and let~$\neg\lambda$ 
be its negation.
Further, let $\widetilde{{\cal C}}[\lambda]$ be the set of concepts of $\lambda$ in $\phi$ and 
let $\widetilde{{\cal C}}[\lambda]^{-}$ be the set of concepts
of type~${\cal C}^{+}$ in~$\widetilde{{\cal C}}[\neg\lambda]$.
It should be clear that $\widetilde{{\cal C}}[\lambda]$ is a subset of~$\widetilde{{\cal C}}$ and
$\widetilde{{\cal C}}[\lambda]^{-}$ is a subset of~$\widetilde{{\cal C}}[\neg\lambda]$.
We say that a set of concepts is of type
$\widetilde{{\cal C}}^{*}$ if all its elements are of type ${\cal C}^{*}$; and a set
of concepts is of type $\widetilde{{\cal C}}^{+}$ if at least one of its elements is of
type ${\cal C}^{+}$. 

We define the understanding $\tilde{u}$ of $\lambda$ with respect to the set $\phi$ as follows:
$$\tilde{u}(\lambda) = \cases{
  \varepsilon, &if $\;$ $\widetilde{{\cal C}}[\lambda]$ is empty $\,$ \hbox{or} 
     $\,$ ($\widetilde{\cal C}[\lambda]^{-}$ is empty \hbox{and} $\widetilde{\cal C}[\lambda]$ 
     is of type $\widetilde{{\cal C}}^{*}$); \cr  
  \noalign{\smallskip}
  t,  &if $\;$ $\widetilde{{\cal C}}[\lambda]$ \hbox{is of type} $\widetilde{{\cal C}}^{+}$
     $\,$ \hbox{and} $\,$ $\widetilde{\cal C}[\lambda]^{-}$ is empty; \cr 
  \noalign{\smallskip}
  f,  &if $\;$ $\widetilde{\cal C}[\lambda]^{-}$ is not empty 
     $\,$ \hbox{and} $\,$ $\widetilde{\cal C}[\lambda]$ is not of type $\widetilde{\cal C}^{+}$.    
     }$$
     
It should be clear that the definition of understanding above leaves one possible case out of 
consideration, that is, if $\widetilde{\cal C}[\lambda]$ is of type $\widetilde{\cal C}^{+}$ 
and $\widetilde{\cal C}[\lambda]^{-}$ is not empty. This is because in such a case, the 
understanding~$\tilde{u}$ is considered {\sl undefined}.

\medskip
\thbegin Lemma A. An understanding~$\tilde{u}$ is defined with respect to a set~$\phi$ of 
clauses \hbox{if and only if} $\tilde{u}$ is equivalent to a satisfying truth assignment 
for~$\phi$.

\proof. We assume first that the understanding~$\tilde{u}$ is equivalent to a satisfying 
truth assignment~$\alpha$ for~$\phi$. It follows that for each clause in~$\phi$ there is 
at least one literal that is true under~$\alpha$. 
Then, based on the equivalence between truth assignments and understandings, defined at 
the beginning of this section, we have that for each clause in~$\phi$ there is at least 
one literal that is true under~$\tilde{u}$ as well.
Therefore $\tilde{u}$ is defined for at least one literal of each of the clauses in~$\phi$, 
and since it is defined to be
true, the rest of literals of each clause must be defined to be free under~$\tilde{u}$. 
Thus, we have that $\tilde{u}$ is defined for all~$\phi$.

For the converse, we show the contrapositive.
Assume we have a truth assignment for all variables in~$\phi$ which is not a satisfying 
truth assignment for~$\phi$, call it~$\overline{\alpha}$. 
This means that in at least one clause of~$\phi$ all literals are assigned to false
under~$\overline{\alpha}$.
Let~$\phi'$ be the set of clauses, subset of~$\phi$, satisfied by~$\overline{\alpha}$ and 
let~$\varphi$ be a clause in~$\phi$ that is not satisfied by~$\overline{\alpha}$ 
(i.e., $\varphi\not\in \phi'$).
From our assumption it follows that $\tilde{u}$ is defined with respect to~$\phi'$ and all 
literals in~$\varphi$ are false under~$\tilde{u}$. 
This implies that $\widetilde{\cal C}[\lambda]^{-}$ is not empty for any literal~$\lambda$ 
in~$\varphi$. Furthermore, the concepts that can be defined of literals 
from~$\varphi$ are all of type~${\cal C}^{+}$.
Therefore, $\tilde{u}$ is not defined for~$\phi$.
\quad$\square$
  
\medskip
Following Lemma~A, for any set~$\phi$ of clauses it might be the case that one or more 
understandings can be defined, or that no understanding can be defined at all. 
We say that two understandings are {\sl equivalent} if both are defined 
with respect to the same set of clauses.

As stated in the introduction, one of the key points of the idea herein presented is that 
based on an understanding defined with respect to a set~$\phi$ of clauses and its 
corresponding set~$\widetilde{\cal C}$ of concepts, it is possible to define all understandings 
that exists with respect to~$\phi$.
In what follows we present two lemmas (Lemma~G and Lemma~D) and two algorithms 
(Algorithm~G and Algorithm~D) that establish some truths and processes
related to the existence of equivalent understandings, which are relevant for our main 
algorithm~(Algorithm~{${\rm \widetilde{U}}$}), presented at the end of this section.

\medskip
\thbegin Lemma G. Let $\tilde{u}$ be an understanding defined with respect to a set $\phi$ 
of clauses, where a literal $\lambda$ in~$\phi$ is free under~$\tilde{u}$.
Let also $\phi_{\lambda}$ be a subset of $\phi$ that contains exclusively all clauses 
from $\phi$ where $\lambda$ or $\neg\lambda$ appear.
If there does not exist an understanding $\tilde{u}_{\lambda}$ defined with respect 
to~$\phi_{\lambda}$ such that $\lambda$ is true under $\tilde{u}_{\lambda}$, then 
there exists no understanding defined with respect to $\phi$ under which $\lambda$ is true.
 
\proof. We let $\widetilde{\cal C}$ be the set of concepts of~$\phi$
from which $\tilde{u}$ is defined. Let also $\widetilde{\cal C}'$ be the set of concepts
that contains only all concepts from $\widetilde{\cal C}$ that correspond to clauses 
in~$\phi_{\lambda}$ (the interpretation of $\widetilde{\cal C}'$ we leave it to 
be $\tilde{u}$ as in~$\widetilde{\cal C}$),   i.e., $\widetilde{\cal C}'$ is a subset 
of~$\widetilde{\cal C}$. 
  
Since we have from $\tilde{u}$ that neither $\lambda$ nor $\neg\lambda$ were needed
for clauses in $\phi_{\lambda}$ to be satisfied (since they are free under $\tilde{u}$),
it follows that an understanding $\tilde{u}_{\lambda}$ under which $\lambda$ is true
can be defined provided that both of the following conditions are met for at least one 
concept~${\cal C}$ in~$\widetilde{\cal C}'[\lambda]$:
\item{\it a)} It is not the case that both of the literals in the definition of ${\cal C}$
are part of the definition of a concept in $\widetilde{\cal C}'[\neg\lambda]$.
\item{\it b)} It is not the case that of the two literals, $l_{1}$ and~$l_{2}$,
that define~${\cal C}$, there is a concept~${\cal C}_{1}$ in~$\widetilde{\cal C}'[\neg\lambda]$
defined by~$l_{1}$ and a literal~$l_{x}$, and there is a concept ${\cal C}_{2}$
in~$\widetilde{\cal C}'[\neg\lambda]$ defined by $l_{2}$ and the negation of~$l_{x}$ 
(i.e., $\neg l_{x}$).
\medskip
From above it follows that the understanding $\tilde{u}_{\lambda}$ cannot exist if no concept in 
$\widetilde{\cal C}'[\lambda]$ meet the conditions stated. And if that is the case, it follows
that there exists no understanding defined with respect to~$\phi$ under which $\lambda$ is true,
because $\phi_{\lambda}$ contains all clauses where $\lambda$ appears, and thus there cannot 
be any concept for~$\lambda$ with respect to $\phi$ that is not already 
in~$\widetilde{\cal C}'[\lambda]$.
   \quad$\square$

\bigskip
\aalgbegin Algorithm G. (Verify if there exists an understanding defined with respect to
a set of clauses that contains exclusively clauses where literals
$\lambda$ or $\neg\lambda$ appear, such that $\lambda$ is true under such understanding).
Given a literal $\lambda$ in a set $\phi$ of clauses, and an understanding $\tilde{u}$
 and set $\widetilde{\cal C}$ of concepts defined with respect to $\phi$,
 such that $\lambda$ is free under $\tilde{u}$.
 Let $\phi_{\lambda}$ be a subset of $\phi$ that contains exclusively all clauses of $\phi$
 where $\lambda$ or $\neg\lambda$ appear. Verify if there exists an understanding
 $\tilde{u}_{\lambda}$ defined with respect to $\phi_{\lambda}$, such that $\lambda$ is true
 under $\tilde{u}_{\lambda}$.
\algstep G1. Set $\tilde{u}'\gets\tilde{u}$ and $\widetilde{\cal C}'\gets \widetilde{\cal C}$.
  Set $\tilde{u}(\lambda)'\gets t$ (this is our assumption) and $\tilde{u}(\neg\lambda)'\gets f$.
\algstep G2. Consider a concept ${\cal C}$ in $\widetilde{\cal C}'[\lambda]$, not yet considered.
  If all concepts in $\widetilde{\cal C}'[\lambda]$ have been considered, the algorithm 
  terminates unsuccessfully; output {\tt False}.
\algstep G3. Let $l_{1}$ and $l_{2}$ be the literals in concept ${\cal C}$. 
  Set both $l_{1}$ and $l_{2}$ to {\sl not true} (i.e., either $\varepsilon$ or $f$) 
  under~$\tilde{u}'$ (following our assumption).
  If $\langle$Compute $\tilde{u}'$$\rangle$ causes no contradiction (in $\tilde{u}'$),
  the algorithm terminates successfully; output {\tt True}. Otherwise, go back to~G2. 
\quad$\blacksquare$

\medskip
The operation $\langle$Compute $\tilde{u}$$\rangle$ used in Algorithm~G above and, later,
in Algorithm~D and Algorithm~{${\rm \widetilde{U}}$}, is defined next.

\bigskip
\halign{#\hfil\cr
$\langle$Compute $\tilde{u}$$\rangle=$ \cr                    
\quad$\;$ Compute $\tilde{u}$ for each literal $\lambda$ and its negation for which the type
of $\widetilde{{\cal C}}[\lambda]$ has changed, \cr  
\quad$\;$  until there is no change of type on any subset of concepts of $\widetilde{\cal C}$. 
  \cr}

\bigskip
\thbegin Lemma D. Let $\tilde{u}$ be an understanding defined with respect to a 
set $\phi$ of clauses, and let $\lambda$ be a literal in $\phi$ that is false 
under~$\tilde{u}$.
Let also ${\cal H}$ be a given set of literals (considered empty, if not given). 
Considering ${\cal H}$, there exists an understanding $\tilde{u}'$ equivalent to~$\tilde{u}$,
such~that $\lambda$ is free under $\tilde{u}'$, \hbox{if and only if}, under 
understanding~$\tilde{u}$, there is at least one literal $l$ (not~in~${\cal H}$) in each 
of the concepts in~$\widetilde{\cal C}[\lambda]^{-}$,
for which the following two conditions are true:
\smallskip
\item{$d_1$.} If $l$ is false under $\tilde{u}$ then,
         considering ${\cal H}'$, defined as ${\cal H}'\gets {\cal H} + \lambda$, 
         there exists an understanding $\tilde{u}''$ equivalent to $\tilde{u}$, 
         such that $l$ is free under $\tilde{u}''$.
\smallskip
\item{$d_2$.}
There exists an understanding, defined with respect to a subset of $\phi$ that contains
exclusively all clauses from $\phi$ where $l$ or $\neg l$ appear, under which $l$ is true. 

\smallskip
\noindent
(Any literal $l$ that is in $\cal H$ is skipped to avoid circular arguments.
It should be clear that if $\cal H$ is empty, the existence of $\tilde{u}'$ is valid
in general; otherwise, the existence or nonexistence of $\tilde{u}'$ 
is only valid for the case in which the elements in $\cal H$ are fixed to be false.)

\proof. 
Let us first assume that $d_1$ and $d_2$ are true for at least one literal $l$
(not in $\cal H$) in each of the concepts in $\widetilde{\cal C}[\lambda]^{-}$.
Condition $d_1$ is true if either $l$ is free under $\tilde{u}$ or
if $l$ is false under $\tilde{u}$ and there exists an understanding
$\tilde{u}''$ defined with respect to $\phi$, such that $l$ is free
under $\tilde{u}''$. Thus, in both cases we have that there
exists an understanding defined with respect to $\phi$ under which $l$ is free.

On the other hand, condition $d_2$ states that there exists an understanding 
defined with respect to a set of clauses $\phi_l$ under which $l$ is true,
where~$\phi_l$ is a subset of $\phi$ that contains exclusively all clauses from $\phi$
where~$l$ or~$\neg l$ appear.
In order for this to be valid not only for $\phi_l$ but for the whole
$\phi$ we need an understanding under which, for at least one concept in 
the set of concepts of $l$, both literals that define it are not true.
One such concept is the one that is obtained from the same clause as 
the concept $\cal C$ in $\widetilde{\cal C}[\lambda]^{-}$. 
Thus, while ${\cal C}$ is defined as $\{\tilde{u}(l), \tilde{u}(l_x)\}$,
the corresponding concept in $\widetilde{\cal C}[l]$ is defined as 
$\{\tilde{u}(\neg\lambda), \tilde{u}(l_x)\}$.
We have then that, under $\tilde{u}$, 
$\neg\lambda$ is true, $\lambda$ is false, and 
$l_x$ is not true (since it is part of a concept in $\widetilde{\cal C}[\lambda]^{-}$).

From above it follows that if we define an understanding $\tilde{u}'$ and related
set $\widetilde{\cal C}'$ of concepts initially as a copy
of $\tilde{u}$ and $\widetilde{\cal C}$ respectively, and
we impose, under $\tilde{u}'$, at least one literal $l$
on each concept in $\widetilde{\cal C}'[\lambda]^{-}$
to be true, then the set $\widetilde{\cal C}'[\lambda]^{-}$ is empty, with which we have
$\lambda$ free under $\tilde{u}'$

\smallskip
For the converse, we show the contrapositive. Thus, we assume that the 
conditions~$d_1$ and~$d_2$ are both, or at least one of them, not true
for at least the two literals~$l_x$ and~$l_y$ that define one of the 
concepts in~$\widetilde{\cal C}[\lambda]^{-}$. 

Let us first assume that $d_1$ is not true. That is, $l_x$ and~$l_y$
are false under~$\tilde{u}$ and there exists no understanding~$\tilde{u}''$
defined with respect to~$\phi$ under which $l_x$ or~$l_y$ is free.
For~$\lambda$ to be free it is necessary that the set of concepts of $\lambda$
is of type $\widetilde{\cal C}^{*}$.
However, based on our assumption one of the concepts of~$\lambda$ is 
defined by two literals ($l_x$ and~$l_y$) that are false under~$\tilde{u}$
and no understanding exists, defined with respect to~$\phi$, under which 
at least of one these literals is free.
This means that $\neg l_x$ and $\neg l_y$ must be true under any 
understanding defined with respect to~$\phi$.
Therefore, the set of concepts of~$\lambda$ is of 
type~$\widetilde{\cal C}^{*}$ under any understanding defined
with respect to~$\phi$.

Finally, we assume that $d_2$ is not true. Let~$\phi_x$ be a subset of clauses 
that contains exclusively all clauses from~$\phi$
where $l_x$ or~$\neg l_x$ appear, and let~$\phi_y$ be a subset of clauses that contains 
exclusively all clauses from~$\phi$ where $l_y$ or~$\neg l_y$ appear.
We assume that there exists no understanding defined with respect to~$\phi_x$
under which $l_x$ is true and there exists no understanding defined with respect to~$\phi_y$
under which $l_y$ is true.
Based on Lemma~G, this assumption implies that there exists no understanding
defined with respect to~$\phi$, such that at least one of $l_x$ or~$l_y$
is true. Therefore, under any understanding defined with respect to~$\phi$
the set of concepts of~$\neg\lambda$ is of type~$\widetilde{\cal C}^{+}$ and
thus $\lambda$ is false.   \quad$\square$

\bigskip
\aalgbegin Algorithm D. (Define an understanding $\tilde{u}'$, equivalent to a given 
understanding~$\tilde{u}$ under which a literal $\lambda$ is false, such that 
$\lambda$ is free under $\tilde{u}'$).
Given an understanding~$\tilde{u}$ and a set~$\widetilde{\cal C}$ of concepts defined
\hbox{with respect to} a set $\phi$ of clauses, a literal $\lambda$ that is false 
under~$\tilde{u}$, and a set ${\cal H}$ of literals (considered empty, if not given).
Define, if possible, an understanding $\tilde{u}'$ and a set $\widetilde{\cal C}'$ 
of concepts equivalent to~$\tilde{u}$ and~$\widetilde{\cal C}$, 
such that $\lambda$ is free under $\tilde{u}'$.
\algstep D0. Set $\tilde{u}'\gets \tilde{u}$ and 
   $\widetilde{{\cal C}}' \gets \widetilde{{\cal C}}$.
\algstep D1. Consider a concept ${\cal C}$ in $\widetilde{\cal C}'[\lambda]^{-}$, 
   not yet considered. 
   If all concepts in $\widetilde{\cal C}'[\lambda]^{-}$ have been considered,
   the algorithms terminates successfully; output~$\tilde{u}'$ and~$\widetilde{\cal C}'$.
\algstep D2. Consider an element $\tilde{u}'(l)$ in ${\cal C}$, not yet considered.
   If all elements of concept ${\cal C}$ have been considered, the algorithm terminates 
   unsuccessfully; output {\tt{\char'140}there is no such understanding{\char'47}}.
\algstep D3. If $l$ is in ${\cal H}$, go back to D2.
\algstep D4. If $l$ is false under $\tilde{u}'$, then 
   set ${\cal H}'\gets {\cal H} + \lambda$ and, based on $\tilde{u}'$ 
   and $\widetilde{\cal C}'$ and considering~${\cal H}'$,
   define if possible an understanding $\tilde{u}''$ and a set $\widetilde{\cal C}''$
   of concepts equivalent to $\tilde{u}'$ and $\widetilde{\cal C}'$,
   such that $l$ is free under $\tilde{u}''$ (this is done by Algorithm D).
   If no such understanding exists, go back to D2.
\algstep D5. If there does not exist an understanding $\tilde{u}_{l}$ defined with respect to
a subset of $\phi$ that contains exclusively all clauses from $\phi$ where $l$ or $\neg l$
appear, such that $l$ is true under $\tilde{u}_{l}$ (checked by Algorithm~G), 
go back to D2.
Otherwise,
if an understanding $\tilde{u}''$ was defined in D4 for $l$, 
set~$\tilde{u}'\gets\tilde{u}''$ and $\widetilde{\cal C}'\gets\widetilde{\cal C}''$,
and, irrespectively of that, 
set~$\tilde{u}'(l)\gets t$ and $\tilde{u}'(\neg l)\gets f$, $\langle$Compute $\tilde{u}'$$\rangle$,
and go back to D1. 
\quad$\blacksquare$

\bigskip
Next, we present our main algorithm which defines for any given 3$\,$SAT problem instance $\Phi$
an understanding with respect to~$\phi$, if one exists, or it determines that none exists.

\bigskip
\aalgbegin Algorithm {${\bf \widetilde{U}}$}. (Define an understanding with respect 
to a 3$\,$SAT problem instance). Given a 3$\,$SAT problem instance $\Phi$, define if possible 
an understanding with respect to $\Phi$.
\algstep {${\bf \widetilde{U}}$}0. Let $\phi$ be an empty set of clauses and let $\tilde{u}$
  be an understanding defined with respect to $\phi$ and $\widetilde{\cal C}$ be an empty set of 
  concepts interpreted according to $\tilde{u}$.
\algstep {${\bf \widetilde{U}}$}1. Consider a clause $\varphi$ that is in $\Phi$
  but not in $\phi$. Assume that $\varphi$ is satisfied. If all clauses in $\Phi$ are in $\phi$,
  the algorithm terminates successfully; $\tilde{u}$ is the answer.
\algstep {${\bf \widetilde{U}}$}2. If all literals in $\varphi$ are false under $\tilde{u}$,
  define if possible an understanding $\tilde{u}'$ and a set $\widetilde{\cal C}'$ of concepts
  equivalent to $\tilde{u}$ and $\widetilde{\cal C}$, such that at least one literal 
  in~$\varphi$ is free under $\tilde{u}'$ (this is done by applying Algorithm~D over each of 
  the literals in $\varphi$ until $\tilde{u}'$ is successfully defined for one of them or all 
  have been processed without success).
  If no understanding $\tilde{u}'$ exists for any of the literals, the algorithm terminates 
  unsuccessfully; output {\tt{\char'140}there exists no understanding with respect 
  to~$\Phi${\char'47}}.
  Otherwise, set $\tilde{u}\gets \tilde{u}'$ and $\widetilde{\cal C}\gets \widetilde{\cal C}'$.
\algstep {${\bf \widetilde{U}}$}3. Consider a literal $\lambda$ in $\varphi$, not yet considered,
  taking first literals that are not false under $\tilde{u}$. If all literals in $\varphi$ have 
  been considered, go back to {${\rm \widetilde{U}}$}1.
\algstep {${\bf \widetilde{U}}$}4. Add the concept of $\lambda$ in $\varphi$ to the 
  set $\widetilde{\cal C}$, $\langle$Compute $\tilde{u}$$\rangle$, and add $\varphi$ to the 
  set $\phi$. \quad$\blacksquare$

\bigskip
Algorithm~{${\rm \widetilde{U}}$} follows straightforward from the definitions and lemmas 
stated previously. One detail, however, is the order in which literals in $\varphi$
are considered in {${\rm \widetilde{U}}$}3, taking first literals that are not false
under $\tilde{u}$. This is meant to avoid getting an undefined understanding form a clause
$\varphi$ where some literals are free, others are false, but none are true under $\tilde{u}$.
Clearly, if the false literals are considered first we will get an undefined understanding. 
However, by taking literals that are not false first we ensure that for one of them its concept
in $\varphi$ will be of type ${\cal C}^{+}$, and for any literal that is false its concept
in $\varphi$ will be of type ${\cal C}^{*}$.

\parindent=15pt
\bigskip\noindent
{\bf 2. Analysis.}\enspace

In this section we present the analysis of our main algorithm (Algorithm~{${\rm \widetilde{U}}$})
in terms of correctness~(\S~2.1) and asymptotic time complexity~(\S~2.2).

\parindent=15pt
\bigskip\noindent
{\bf 2.1 Correctness.}\enspace

\medskip
\thbegin Theorem 1. Algorithm~$\rm\widetilde{U}$ terminates successfully 
\hbox{if and only if} $\Phi$ is satisfiable. 

\medskip
We prove Theorem~1 through a sequence of lemmas.

\medskip
\thbegin Lemma 1. If $\Phi$ is satisfiable, Algorithm~$\rm\widetilde{U}$ terminates successfully.
\proof. 
We assume initially that $\Phi$ is satisfiable. 

The proof is by induction, where the induction hypothesis is that there exists an 
understanding defined with respect to a set $\phi$ of clauses, subset of $\Phi$.

The base case is for~$|\phi|=1$. In this case an understanding~$\tilde{u}$
is always defined with respect to~$\phi$, with the first literal considered
in~{${\rm \widetilde{U}}$}3 made true in~{${\rm \widetilde{U}}$}4,
since its concept is of type~${\cal C}^{+}$ (due to literals in that concept being
initially free), and then the type of the concept of the other two literals 
is~${\cal C}^{*}$, making them free under~$\tilde{u}$.

For the induction step, we wish to show that there exists an understanding
defined with respect to~$\phi+\varphi$, where~$\phi$ is a subset of~$\Phi$
and $\varphi$ is a clause in~$\Phi$ but not in~$\phi$.
In all but one case there exists an understanding defined with respect 
to~$\phi+\varphi$. Such case happens if all literals in~$\varphi$ are false
under all understandings that can be defined with respect to~$\phi$.
Based on the induction hypothesis, there exists an understanding
defined with respect to~$\phi$. And, from our initial assumption (i.e., 
$\Phi$ is satisfiable) we have that there exists a truth assignment which 
satisfies all clauses, thus there is no clause in $\Phi$ with all its literals
assigned to false. 
Therefore, the call to Algorithm~D in~{${\rm \widetilde{U}}$}2, defines
successfully an understanding with respect to~$\phi$ under which one 
literal~$\lambda$ in~$\phi$ is free. Then, in~{${\rm \widetilde{U}}$}3,
$\lambda$ is considered first since the other two literals in~$\varphi$
are false. And finally in~{${\rm \widetilde{U}}$}4, understanding~$\tilde{u}$
is defined with respect to~$\phi+\varphi$.  \quad$\square$

\medskip
\thbegin Lemma 2. If Algorithm~$\rm\widetilde{U}$ terminates successfully then $\Phi$ 
is satisfiable.
\proof. We show the contrapositive. Thus, we assume that $\Phi$ is not satisfiable.
Based on Lemma~A we use the equivalence. That is, our assumption is that there exists 
no understanding defined with respect to $\Phi$.

For our assumption to be true it is necessary that for at least one clause~$\varphi$,
in~$\Phi$, all its literals are false under all understandings that can be defined
with respect to a set~$\phi$, subset of~$\Phi$, that does not include~$\varphi$.
In such a case, Algorithm~{${\rm \widetilde{U}}$} executes~{${\rm \widetilde{U}}$}2,
where Algorithm~D is executed to try to define an understanding with respect to~$\phi$
under which one of the literals in $\varphi$ is free. However, due to our assumption 
Algorithm~D fails. Consequently, Algorithm~{${\rm \widetilde{U}}$}
terminates unsuccessfully.  \quad$\square$

\medskip
\noindent
This concludes the proof of Theorem 1.

\parindent=15pt
\bigskip\noindent
{\bf 2.2 Time Complexity.}\enspace

\bigskip
\thbegin Theorem 2. For any given 3$\,$SAT problem instance $\Phi$, 
Algorithm~$\rm\widetilde{U}$ terminates in polynomial time.
\proof. We analyze the algorithm complexity in two parts.

The first part is concerned with {${\rm \widetilde{U}}$}1, {${\rm \widetilde{U}}$}3, 
and {${\rm \widetilde{U}}$}4. These steps perform a constant number of operations
on the number of literals in~$\varphi$, except for~$\langle$Compute $\tilde{u}$$\rangle$
which, in case the type of the set of concepts of~$\lambda$ has changed,
it has to recompute~$\tilde{u}$ for~$\lambda$ and its negation and check if it is necessary
to recompute~$\tilde{u}$ for any other literal for which $\lambda$
is part of the definition of its set of concepts.
In the worst case this process goes through all concepts that have been defined
with respect to~$\phi$. That is, at most three times the number of clauses in~$\phi$.
Thus, 
if we consider that at every iteration Algorithm~{${\rm \widetilde{U}}$}
should go through this worst case (until all clauses in~$\Phi$ are processed),
we get roughly an arithmetic series as the number of operations performed.

The second part is concerned with~{${\rm \widetilde{U}}$}2, where assuming
that all literals in~$\varphi$ are false, Algorithm~D will be executed for
each literal~$\lambda$ in~$\varphi$ until it defines an understanding under 
which $\lambda$ is free. 
The number of iterations in Algorithm~D depends on the number of concepts
in~$\widetilde{\cal C}[\lambda]^{-}$, for the literal~$\lambda$ for which
Algorithm~D is meant to define an understanding.
We recall that $\widetilde{\cal C}[\lambda]^{-}$ is the set of concepts
of type~${\cal C}^{+}$ of~$\neg\lambda$.
Since by definition there can be only one concept of type~${\cal C}^{+}$
defined from each clause, we have that the number of concepts of 
type~${\cal C}^{+}$ in the set of concepts defined with respect 
to~$\phi$ is at most equal to the number of clauses in~$\phi$.
Thus, the maximum number of iterations of Algorithm~D overall
(including its recursive call in~D4 for some literals in concepts
in~$\widetilde{\cal C}[\lambda]^{-}$)
is bounded by the total number of clauses in~$\phi$~(times some constant).
Therefore, if we consider that at every iteration
Algorithm~{${\rm \widetilde{U}}$} should execute Algorithm~D
in {${\rm \widetilde{U}}$}2 over each of the literals of clause~$\varphi$,
we have in the worst case roughly an arithmetic series as the total number
of operations. 

In both parts above we have an upper bound of approximately $O(m^{2})$, where 
$m$ is the number of clauses in $\Phi$. Therefore, 
Algorithm~{${\rm \widetilde{U}}$} terminates in polynomial time.
\quad$\square$

}

\bigskip
\centerline{\bf Bibliography}
\bigskip

{\advance\baselineskip -1pt \parskip5pt plus .5pt minus .5pt

\def\bib[#1] {\par\noindent\hangindent\parindent\hbox to\parindent{[#1]\hfil}}

\smallskip
\bib
[\Cook]
S. A. Cook, ``The complexity of theorem-proving procedures,'' In {\sl
Proceedings of the third annual ACM symposium on Theory of computing\/},
pages 151--158. ACM, 1971.

\smallskip
\bib
[\Levin]
L. A. Levin, ``Universalnye zadachi perebora''  [Universal sequential search problems], {\sl Problemy
Peredachi Informatsii\/}, 9(3), pages 265--266, 1973. 
English translation in: B. A. Trakhtenbrot. A survey of russian
approaches to Perebor (brute-force search) algorithms. Annals of the History of
Computing, 6(4), pages 384--400, 1984.

\smallskip
\bib
[\Karp]
R. M. Karp, ``Reducibility among combinatorial problems,'' In R. Miller, J. Thatcher,
and J. Bohlinger, editors, {\sl Complexity of Computer Computations\/}, The IBM Research
Symposia Series, pages 85--103. Springer US, 1972.

\smallskip
\bib
[\Hema]
L. A. Hemaspaandra, ``Sigact news complexity theory column 36,''
{\sl ACM SIGACT News\/}, 33(2), pages 34--47, 2002.

\smallskip
\bib
[\Gasa]
W. I. Gasarch, ``Guest column: The second P=? NP poll,''
{\sl ACM SIGACT News}, 43(2), pages 53--77, 2012.


\smallskip
\bib
[\Davis]
M. Davis and H. Putnam, ``A computing procedure for quantification theory,'' 
{\sl Journal of the ACM (JACM)\/}, 7(3), pages 201--215, 1960.

\smallskip
\bib
[\Loge]
M. Davis, G. Logemann, and D. Loveland, ``A machine program for theorem-proving,'' 
{\sl Communications of the ACM\/}, 5(7), pages 394--397, 1962.

\smallskip
\bib
[\Marques]
J. P. Marques-Silva and K. A. Sakallah, ``GRASP--a new search algorithm for satisfiability,'' 
In {\sl Proceedings of the 1996 IEEE/ACM international conference on Computer-aided design\/}, 
pages 220--227. IEEE, 1997.

\smallskip
\bib
[\Selman]
B. Selman, H. Levesque, and D. Mitchell, ``A New Method for Solving Hard Satisfiability Problems,''
In {\sl Proceedings of the Tenth National Conference on Artificial Intelligence (AAAI'92)\/}, 
pages 440--446. AAAI, 1992.

\smallskip
\bib
[\Garey]
M. R. Garey and D. S. Johnson, {\sl Computers and intractability, A guide to the theory of NP-Completeness} 
(New York: W. H. Freeman,1979).


}

\bye